\documentclass[10pt,twoside]{article}
\usepackage{amsmath,amsfonts,amssymb}
\usepackage{latexsym}
\usepackage{dcolumn}
\usepackage{graphicx,epsfig}
\usepackage{amsthm}
\usepackage{hyperref}

\newcommand{\ud}{\,\mathrm{d}}

\begin{document}

\begin{flushright}
{\tt IISER(Kolkata)/GR-QC\\ \today}
\end{flushright}
\vspace{1.5cm}

\begin{center}
{\Large \bf Dilaton cosmology and the Modified Uncertainty Principle}
\vglue 0.5cm
Barun Majumder\footnote{barunbasanta@iiserkol.ac.in}
\vglue 0.6cm
{\small {\it Department of Physical Sciences,\\Indian Institute of Science Education and Research (Kolkata),\\
Mohanpur, Nadia, West Bengal, Pin 741252,\\India}}
\end{center}
\vspace{.1cm}

\begin{abstract} 
Very recently Ali $\textit{et al.}$(2009) proposed a new generalized uncertainty principle (or GUP) with a linear term in Plank length which is consistent with
doubly special relativity and string theory. The classical and quantum effects of this generalized uncertainty principle (term as modified uncertainty principle or MUP) are investigated on the phase space of a dilatonic cosmological model with an exponential dilaton potential in a flat FRW background . Interestingly, as a consequence of MUP, we found that it is possible to get a late time acceleration for this model. For the quantum mechanical description in both commutative and MUP framework, we found the analytical solutions of the Wheeler-DeWitt equation for the early universe and compare our results. We have used an approximation
method in the case of MUP. 
\vspace{5mm}\newline Keywords: dilaton, quantum cosmology, generalized uncertainty principle
\end{abstract}
\vspace{.3cm}

\section{Introduction}

\hspace{.5cm} Predictions from any candidate theory of quantum gravity can be well tested by cosmology as this area of physics can test physics at energies much higher
than the collider experiments. Until we get a satisfactory theory of quantum gravity we can study the applications of the existing theories in 
cosmological scenarios. Quite recently string theory has made remarkable efforts to solve some of the age old problems in understanding our
universe \cite{z1,z2,z3,z4}. In the pre big bang scenario based on the string effective action \cite{z5}, the birth of the universe is described by a
transition from the string perturbative vacuum with weak coupling, low curvature and cold state to the standard radiation dominated regime, passing through a
high curvature and strong coupling phase. This transition is made by the kinetic energy term of the dilaton, a scalar field with which the Einstein-Hilbert
action of general relativity is augmented (also see \cite{z6} and the references therein for a brief review of string dilaton cosmology). Duality is one of
the key feature of string dilaton cosmology. If the scale factor is a solution of the equation of motion then the inverse of the former is also a solution. This
indicates that our universe behaves like a string, i.e., has a minimal size of the order of string scale and maximal size of the order of inverse string scale.
\par
The idea that the uncertainty principle could be affected by gravity was first given by Mead \cite{z7}. Later modified commutation relations between position
and momenta commonly known as the generalized uncertainty principle (or GUP) were proposed by candidate theories of quantum gravity such as, string
theory, doubly special relativity theory (or DSR) and black hole physics with the prediction of a minimum measurable length \cite{z8,ext1,z9,z10}. Similar kind
of commutation relation can also be found in the context of polymer quantization in terms of polymer mass scale \cite{z11}. From the point of view of
perturbative string theory, strings cannot probe distances smaller than their own size. This is the key idea for the prediction of a minimum measurable
length. To incorporate the existence of a minimal length in usual quantum mechanics, the standard commutation relation between position and momenta gets
modified along with the Hilbert space representation \cite{z9}. In one dimension the form of this generalized uncertainty principle can be expressed
by the associative Heisenberg algebra generated by $x$ and $p$ obeying the relation ($\beta>0$)
\begin{equation}
[x,p] = i~ \hbar ~ (1+\beta ~p^2)~~.
\end{equation}
The corresponding uncertainty relation can be written as
\begin{equation}
\Delta x ~\Delta p \geq \frac{\hbar}{2}~ [ ~1 + \beta ~(\Delta p)^2 + \beta ~\langle p \rangle ^2 ~]~~. 
\end{equation}
It is clear from the above equations that the smallest uncertainty in position has the value
\begin{equation}
\Delta x_{min}~= \hbar ~ \sqrt{\beta}~~.
\end{equation}
If $\beta=0$ we get usual Heisenberg uncertainty relation. Here $\beta$ must be related to the Plank length. More general cases and applications of such
commutation relations are studied in \cite{z12}. Since the scale factors, matter fields and their conjugate momenta play the role of dynamical variables
in cosmological models so the application of the generalized uncertainty principle in constructing the phase space of these models is reasonably relevant. Many
researchers have expressed a vested interest in studying the classical and quantum solutions of some cosmological models in the GUP
framework and also in noncommutative literature \cite{z13,z14,zzzz1}. 
\par
The authors in \cite{z15} proposed a GUP which is consistent with DSR theory, string theory and black hole physics and which says
\begin{equation}
\left[x_i,x_j\right] = \left[p_i,p_j\right] = 0 ,
\end{equation}
\begin{equation}
\label{e2}
[x_i, p_j] = i \hbar \left[  \delta_{ij} -  l  \left( p \delta_{ij} +
\frac{p_i p_j}{p} \right) + l^2  \left( p^2 \delta_{ij}  + 3 p_{i} p_{j} \right)  \right],
\end{equation}
\begin{align}
\label{e3}
 \Delta x \Delta p &\geq \frac{\hbar}{2} \left[ 1 - 2 l \langle p\rangle + 4 l^2 \langle p^2\rangle \right]  \nonumber \\
& \geq \frac{\hbar}{2} \left[ 1  +  \left(\frac{l}{\sqrt{\langle p^2 \rangle}} + 4 l^2  \right)  \Delta p^2  +  4 l^2 \langle p \rangle^2 -  2 l \sqrt{\langle p^2 \rangle} \right],
\end{align}
where $p^2=\sum_{k=1}^3 p_k p_k$, $x^2=\sum_{k=1}^3 x_k x_k$ and $ l=\frac{l_0 l_{pl}}{\hbar} $. Here $ l_{pl} $ is the Plank length ($ \approx 10^{-35} m $). For a brief discussion on equation (\ref{e2}) see
Appendix. It is normally assumed that the dimensionless
parameter $l_0$ is of the order unity. If this is the case then the $l$ dependent terms are only important at or near the Plank
regime. But here we expect the existence of a new intermediate physical length scale of the order of $l \hbar = l_0 l_{pl}$. We also note
that this unobserved length scale cannot exceed the electroweak length scale \cite{z15} which implies $l_0 \leq 10^{17}$. These equations are
approximately covariant under DSR transformations but not Lorentz covariant \cite{z10}. These equations also imply
\begin{equation}
\Delta x \geq \left(\Delta x \right)_{min} \approx l_0\,l_{pl}
\end{equation}
and
\begin{equation}
\Delta p \leq \left(\Delta p \right)_{max} \approx \frac{M_{pl}c}{l_0}
\end{equation}
where $ M_{pl} $ is the Plank mass and $c$ is the velocity of light in vacuum. It can be shown that equation (\ref{e2}) is satisfied by the
following definitions (see Appendix)
\begin{equation}
\label{defi}
x_i=x_{oi} ~~\text{and}~~ p_i=p_{oi} (1 - l\,p_o + 2\,l^2\,p_o^2) ~~,
\end{equation}
where $x_{oi}$, $p_{oj}$ satisfies $[x_{oi}, p_{oj}]= i \hbar \delta_{ij}$. Here we can interpret $p_{oi}$ as the momentum at low energies having the standard representation in position space ($ p_{oi}\equiv -i\hbar \frac{\partial}{\partial x_{oi}}$) with $p_o^2=\sum_{i=1}^3 p_{oi}p_{oi}$ and $ p_i $ as the momentum at high energies. Using (\ref{defi}) (to the best of our knowledge, this definition is unique) we can also show that the $ p^2 $ term in the kinetic part of any Hamiltonian (including relativistic ones) can be written as \cite{z15}
\begin{equation}
\label{po}
p^2 \Longrightarrow \ p_o^2 - 2\ l\ p_o^3 + {\cal O}(l^2) + \ldots ~~~ .
\end{equation}
Here we neglect terms $ {\cal O}(l^2)$ and higher in comparison to terms $ {\cal O}(l)$ to study the effect of the linear term in $l$ in the first approximation as $l=l_0\,l_{pl}$. Given the robust nature of GUP, such corrections will continue to play a role irrespective of what other quantum gravity corrections one
may consider. In other words, they are in some sense universal! The effect of this proposed GUP is well studied recently for some well known physical systems in \cite{z15,z16}.
\par
In this paper we are going to study a dilaton cosmological model with an exponential dilaton potential. We will study the model with a suitable metric and a particular lapse function. At the beginning we will concentrate on the classical solutions. Later we will study the effect of the modified uncertainty
principle (from now on we will use the terminology Modified Uncertainty Principle or MUP for the recently proposed GUP \cite{z15} to distinguish it from
the earlier version) on the classical model. In the next half of the paper we are going to study the quantum model. In each case we are going to solve
the Wheeler-DeWitt equation but with a relevant approximation and compare the results. As $l_{pl}=\sqrt{\hbar G/c^3}$, where $G$ is the Newtonian coupling constant, we can imply that the extra terms in the uncertainty relation (\ref{e2}) is a consequence of strong gravity. So studying early universe quantum cosmological models in the MUP framework is physically relevant. Here we will apply the minisuperspace approach of quantum gravity where one reduces a large number of degrees of freedom by imposing symmetry conditions on the metric. Since our model has two degrees of freedom, the scale factor and the dilaton
field, the study of this model within the MUP framework appears to have physical grounds.

\section{The Dilaton Model}

\hspace{.5cm} The four dimensional gravi-dilaton effective action in the string frame can be written as \cite{z17}
\begin{equation}
\label{a1}
{\cal S} = - \frac{1}{2\lambda_s} \int \ud^4 x \sqrt{-g}~ e^{-\phi} ~[ ~{\cal R} + \partial_{\mu}\phi ~\partial^{\mu}\phi + V(\phi)~]~~,
\end{equation}
where $\phi$ is the dilaton field, $\lambda_s$ is the fundamental string length ($l_s$) parameter and $V(\phi)$ is the dilaton potential. In the string frame the fundamental unit is the string length $l_s$, and thus the Planck mass, which is the effective coefficient of the Ricci scalar ${\cal R}$, varies with the
dilaton. An alternative way to describe things is to use the Einstein frame which is physically more transparent than the string frame. In the Einstein
frame it is the Plank length which is more directly related to the macroscopic physics through the strength of gravity which is used as a fundamental unit. So we would like to prefer the Einstein frame and in this frame the action (\ref{a1}) takes the form \cite{z4}
\begin{equation}
{\cal S} = -\frac{M^2}{2} \int \ud^4 x \sqrt{-g}~ [ ~{\cal R} - \frac{1}{2}~ \partial_{\mu}\phi ~\partial^{\mu}\phi - V(\phi) ~]~~,
\end{equation} 
where $M$ is the four dimensional Plank mass. It is to be noted that now the sign of the kinetic term of the scalar field is familiar to us. We consider a
spatially flat Friedmann-Robertson-Walker spacetime which is specified by the metric of the form \cite{z18}
\begin{equation}
\label{metric}
\ud s^2 = -\frac{N^2(t)}{a^2(t)} \ud t^2 + a^2(t)~ \delta_{ij} \ud x^i \ud x^j~~,
\end{equation} 
where $N(t)$ is the lapse function and $a(t)$ is the scale factor of the universe. The effective Lagrangian of the model can be easily expressed as
\begin{equation}
\label{lagran}
{\cal L} = \frac{1}{N} \left( -\frac{1}{2} a^2 \dot{a}^2 + \frac{1}{2} a^4 \dot{\phi}^2 \right) -N a^2 V(\phi)~~.
\end{equation}
We now use a set of transformations
\begin{equation}
\label{trans}
x = \frac{a^2}{2} \cosh(\alpha \phi) ~~~~, ~~~~y = \frac{a^2}{2} \sinh(\alpha \phi) ~~,
\end{equation}
where $\alpha$ is a positive constant, so that we can finally write the Hamiltonian decoupled in the variables $x$ and $y$. In these new variables
the Lagrangian can be written as
\begin{equation}
{\cal L} = \frac{1}{2N} ~(\dot{y}^2 - \dot{x}^2 )- 2N~(x-y) ~e^{\alpha \phi}~ V(\phi)~~. 
\end{equation}
Here we take the opportunity to choose the potential as
\begin{equation}
V(\phi) = \frac{V_0}{2}~e^{-\alpha \phi}
\end{equation}
so that the Lagrangian can be written as
\begin{equation}
{\cal L} = \frac{1}{2N}~(\dot{y}^2 - \dot{x}^2) - N V_0~ (x-y)~~. 
\end{equation}
We have a physical motivation for choosing this potential as this type of potential finds its use in quintessence model of dark
energy (for a brief review see \cite{z19} and the references therein) and also used in an inflationary model \cite{z20}. With the above Lagrangian the
Hamiltonian constraint can be written as 
\footnote{In general, the main reason for applying the transformation like (\ref{trans}) is that the minisuperspace is curved in terms of the original coordinates (see (\ref{lagran})) and the application of some deformed commutation relation like GUP is not an easy task. However, we know that any 2-D Riemannian
space is conformally flat and if the Ricci scalar is zero this space will be Minkowskian. This means that there is a set of coordinates in terms of
which the metric is flat. In our case the 2-D minisuperspace (spanned by $a$ and $\phi$) is a Riemannian space with vanishing Ricci scalar. Therefore one can find a
coordinate transformation (\ref{trans}) in terms of which the minisupermetric takes the form of a Minkowskian space (see(\ref{ham})). We also note that we did not introduce a canonical transformation and therefore after introducing the new coordinates all of the quantum analysis should be done based on them and
one should not make an inverse transformation in the wave functions.}
\begin{equation}
\label{ham}
{\cal H} = -\frac{1}{2}~ p_x^2 + \frac{1}{2} ~p_y^2 + V_0 ~(x-y)~~.
\end{equation}
The minisuperspace of this model is a two dimensional manifold $0<a<\infty$, $-\infty<\phi<\infty$. Its non-singular boundary is the
line $a=0$ with $\vert \phi \vert<\infty$, while at singular boundaries at least one of the two variables is infinite \cite{z21}. In terms of the new
variables $x$ and $y$ the minisuperspace is recovered by $x>0$, $x>\vert y\vert$ and the non singular boundary may be represented by $x=y=0$ \cite{z14}.

\section{Classical Dilaton Cosmology}

\hspace{.5cm} In this section we are going to study the classical solutions for the above Hamiltonian. We will study the commutative case followed by
the MUP effects on the model.

\subsection*{Commutative case}

In the gauge $N=1$ we can get the equations of motion for the above Hamiltonian keeping in mind that the phase space variables satisfy
\begin{equation}
\{x_i,x_j\} = \{p_i,p_j\} = 0 ~~~~,~~~~ \{x_i,p_j\} = \delta_{ij} ~~.
\end{equation}
The equations of motion for the Hamiltonian in (\ref{ham}) are 
\begin{align}
& \dot{x} = \{x,{\cal H}\} = -p_x ~~~~,~~~~ \dot{p_x} = \{p_x,{\cal H}\} = -V_0 ~~,\nonumber \\
& \dot{y} = \{y,{\cal H}\} = p_y ~~~~ \text{and} ~~~~ \dot{p_y} = \{p_y,{\cal H}\} = V_0 ~~.
\end{align}
A straight forward integration yields
\begin{align}
& x(t) = \frac{1}{2} V_0 t^2 - p_{ox}t + x_0 ~~~~,~~~~ p_x(t) = -V_0t + p_{0x} ~~,\nonumber \\
& y(t) = \frac{1}{2} V_0 t^2 + p_{oy}t + y_0 ~~~~ \text{and} ~~~~ p_y(t) = V_0t + p_{0y} ~~.
\end{align}
Here $x_0,y_0,p_{0x},p_{0y}$ are the arbitrary integration constants to be evaluated with suitable initial conditions. With these solutions and the
constraint equation ${\cal H}=0$ it can be easily shown that classically only half of the minisuperspace $x>y>0$ (or $a>0, \phi>0$) is recovered by the dynamical variables $x(t)$ and $y(t)$. Considering $x_0=y_0$ and $p_{0x}=p_{0y}$ and using the transformation (\ref{trans}) we can get
\begin{equation}
\label{scale}
a(t) = \left(~8~\vert p_{0x}\vert ~ V_0 ~t^3 + 16~ x_0 ~\vert p_{0x}\vert ~t~\right)^{\frac{1}{4}}
\end{equation}
and
\begin{equation}
\label{dil}
\phi(t) = \frac{1}{2\alpha} \ln \left(\frac{V_0~t^2 + 2x_0}{2~\vert p_{0x} \vert ~t}\right) ~~.
\end{equation}
\begin{figure}[htb]
\begin{tabular}{c}
\hspace{2.5cm} \includegraphics[width=10cm,height=6.5cm]{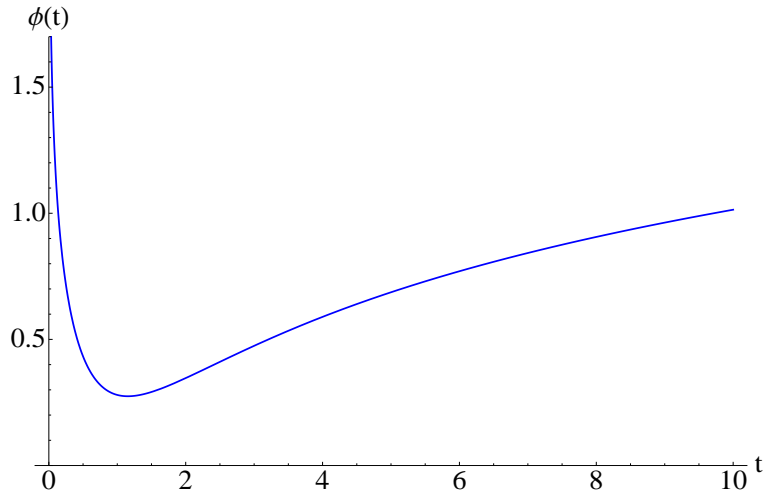}
\end{tabular}
\caption{\footnotesize  The classical behaviour of $\phi$ with respect to time. The plot is with numerical
values $\alpha=1,V_0=1.5,x_0=1$ and $\vert p_{0x} \vert=1$. We have checked the limit of $\phi$ in the limit $t\rightarrow 0$ and it diverges.}
\label{1phi}
\end{figure}
In fig.(\ref{1phi}) we have shown the time evolution of the dilaton field. The approximate time dependence of the scale factor and the field at earlier times
can be written as
\begin{equation}
\label{early1}
a(t) \sim t^{\frac{1}{4}} ~~~~ \text{and} ~~~~ \phi(t) \sim \ln \left(\frac{1}{t}\right) ~~,
\end{equation}
and the late time behaviour can be written as
\begin{equation}
\label{late1}
a(t) \sim t^{\frac{3}{4}} ~~~~ \text{and} ~~~~ \phi(t) \sim \ln (t) ~~.
\end{equation}
To get the standard parametrization of the FRW metric one should make the gauge choice $N(t)=a(t)$ because in that gauge the resultant time is the cosmic
time (see equation (\ref{metric})). But here we made the choice $N=1$. So we have to translate the time of equation (\ref{scale}) and (\ref{dil}) in to
cosmic time with the gauge choice
\begin{equation}
\ud \tau = \frac{1}{a(t)} \ud t ~~,
\end{equation}
where $\tau$ represents the cosmic time. With this choice equation (\ref{early1}) and (\ref{late1}) can be rewritten as
\begin{equation}
a(\tau) \sim \tau^{\frac{1}{3}} ~~~~ \text{and} ~~~~ \phi(t) \sim \ln \left(\frac{1}{\tau}\right) ~~,
\end{equation}
and the late time behaviour can be written as
\begin{equation}
\label{acc}
a(\tau) \sim \tau^3 ~~~~ \text{and} ~~~~ \phi(\tau) \sim \ln (\tau) ~~.
\end{equation}
Equation (\ref{acc}) reminds us about the well known fact that if $a \sim (\text{Time})^n$, the universe will have an accelerated expansion or a decelerated
expansion for $n>1$ or $n<1$ respectively throughout the period (for e.g., see \cite{z22}). So we get a decelerated expansion at earlier times and an accelerated expansion at late times.
Our result contradict the classical result of \cite{z14} for the time dependence of the dilaton field at an earlier time but agrees quite well with the classical
result of \cite{z23}. Our result is physically more acceptable, at least from the classical point of view.

\subsection*{Classical Dilaton Cosmology in the MUP framework}

\begin{figure}[htb]
\begin{tabular}{c}
\hspace{2.5cm} \includegraphics[width=10cm,height=5.5cm]{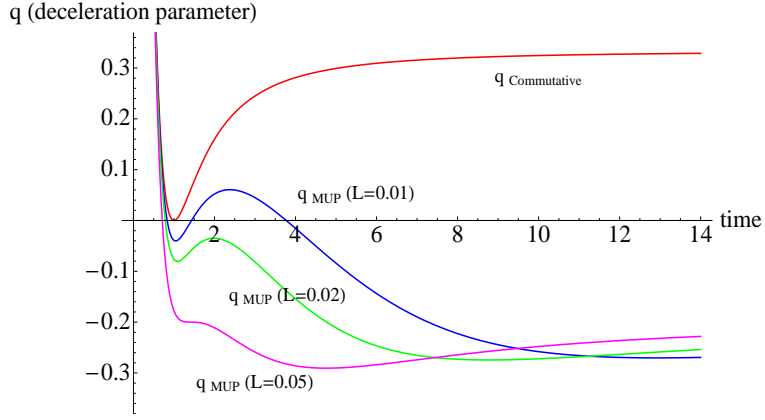}
\end{tabular}
\caption{\footnotesize  Plot of the deceleration parameter ($q$) with time in the commutative case and in the MUP framework. Here we have considered the
numerical values of the constants in such a way that $(8 ~\vert p_0 \vert ~c + 12 ~l ~c ~\vert p_0 \vert^2 ~)=1 ~\text{and} ~( 4~ \vert p_0 \vert ~V_0 + 18 ~l ~V_0 ~\vert p_0 \vert^2 +  16 ~l ~V_0^2 ~c )=1$. We have defined $L=8~l~V_0^3$ and we give the plots for three different values of $L$.}
\label{2figacc}
\end{figure}

\hspace{.5cm} The motivation for studying this cosmological model in the MUP framework lies in the fact that we have a linear term in Plank length
in the commutation relation (see equation (\ref{e2}) and (\ref{e3})). Here we would like to explore the consequences of this linear term in
Plank length (as $l=l_0l_{pl}$ with $\hbar=1$) and compare our results with the commutative case. Following equation (\ref{po}) we rewrite equation (\ref{ham}) as
\begin{equation}
\label{hmup}
{\cal H}_{MUP} = -\frac{p_1^2}{2} + l ~p_1^3 + \frac{p_2^2}{2} - l~ p_2^3 + V_0 ~(x-y) + {\cal O}(l^2) + \ldots ~~,
\end{equation}
where $p_{1,2}$ can be interpretated as the momentum at low energies having the standard representation in position
space (i.e., $p_{1,2}\equiv -i \frac{\partial}{\partial x,y}$). Throughout this paper we will neglect terms ${\cal O}(l^2)$ and higher to study the
linear effect of $l$ perturbatively. The equations of motion in this case are
\begin{equation}
\label{mot1}
\dot{x} = \{x,{\cal H}_{MUP}\} = -p_1  + 3~l~p_1^2 ~~~~ \text{and} ~~~~ \dot{p_1} = \{p_1,{\cal H}_{MUP}\} = -V_0 ~~, 
\end{equation}
\begin{equation}
\label{mot2}
\dot{y} = \{y,{\cal H}_{MUP}\} = p_2  - 3~l~p_2^2 ~~~~ \text{and} ~~~~ \dot{p_2} = \{p_2,{\cal H}_{MUP}\} = V_0 ~~.
\end{equation}
Adding the second equations of (\ref{mot1}) and (\ref{mot2}) we get
\begin{equation}
\dot{p_1} + \dot{p_2} = 0 \rightarrow p_1 + p_2 = p_0~ (\text{a const.}) ~~.
\end{equation}
The solution for $p_1$ and $p_2$ can be written after integration as
\begin{equation}
\label{p12}
p_1 (t) = -V_0~t + \frac{p_0}{2} ~~~~ \text{and} ~~~~ p_2 (t) = V_0~t + \frac{p_0}{2} ~~.
\end{equation}
With equation (\ref{p12}) we can easily integrate the first equations of (\ref{mot1}) and (\ref{mot2}) for the solution
\begin{equation}
x(t) = \left(-\frac{p_0}{2} + \frac{3~l~p_0^2}{4}\right)~t + \left(\frac{V_0}{2} - \frac{3~l~V_0~p_0}{2}\right)~t^2 + l~V_0^2~t^3 + x_0 ~~,
\end{equation}
and
\begin{equation}
y(t) = \left(\frac{p_0}{2} - \frac{3~l~p_0^2}{4}\right)~t + \left(\frac{V_0}{2} - \frac{3~l~V_0~p_0}{2}\right)~t^2 - l~V_0^2~t^3 + y_0 ~~.
\end{equation}
Now these solutions must satisfy the zero energy condition $H_{MUP}=0$. For that we require $x_0=y_0=c ~(\text{a const.})$. Using the
transformation (\ref{trans}) we can calculate the time dependence of the scale factor and the dilaton field as 
\begin{align}
\label{scalemup}
a(t) =& ~[ ~\{8 ~\vert p_0 \vert ~c + 12 ~l ~c ~\vert p_0 \vert^2 ~\}~t + \{ 4~ \vert p_0 \vert ~V_0 + 18 ~l ~V_0 ~\vert p_0 \vert^2 +  16 ~l ~V_0^2 ~c \}~t^3 \nonumber \\
   &+ 8 ~l ~V_0^3 ~t^5 ~ ]^{\frac{1}{4}} ~~,
\end{align}
and
\begin{align}
\phi(t) =& \frac{1}{\alpha}\Big[ \ln ~\{ (8~\vert p_0 \vert ~c + 12 ~l ~c ~\vert p_0 \vert^2 )~ t + ( 4 ~\vert p_0 \vert ~V_0 + 18 ~l ~V_0 ~\vert p_0 \vert^2 +
16 ~l ~V_0^2 ~c )~ t^3  \nonumber \\
&+ 8 ~l ~V_0^3 ~t^5~ \}^{\frac{1}{2}} - \ln ~\{ (2 ~\vert p_0 \vert + 3 ~l ~\vert p_0 \vert^2 ) ~t + 4 ~l ~V_0^2 ~t^3 \} \Big] ~~.
\end{align}
Here we have neglected terms ${\cal O}(l^2)$. If we set $l=0$ we get back equations (\ref{scale}) and (\ref{dil}). At earlier times the limiting
behaviour of $a(t)$ and $\phi(t)$ can be written as
\begin{equation}
a(t) \sim t^{\frac{1}{4}} ~~~~ \text{and} ~~~~ \phi(t) \sim \ln(t) ~~.
\end{equation}
Researchers from the field of $\textit{dark energy}$ define a parameter known as deceleration parameter ($q$) and it is written as (for e.g., see \cite{z22})
\begin{equation}
q = -~\frac{a ~\ddot{a}}{\dot{a}^2} ~~.
\end{equation}
In cosmology this is a dimensionless measure of the cosmic acceleration. From the recent observations we know that our universe is expanding in an accelerated
rate. For that we require $q$ to be negative. For equation (\ref{scale}) and equation (\ref{scalemup}) we have plotted the value of this deceleration
parameter with respect to time in fig.(\ref{2figacc}). Fixing the constants $l,\vert p_0 \vert ,V_0 ~\text{and} ~c$ to some values, it is possible to get a late
time acceleration in this MUP framework of the model. 
So from fig.(\ref{2figacc}) we can clearly see that the modified uncertainty principle not only has its effects on the early universe cosmology but it is also possible that it can provide an accelerated expansion of the universe at late times.

\section{Quantum Dilaton Cosmology}

\hspace{.5cm} Here we are going to write the Wheeler-DeWitt equation for the Hamiltonian constraint ${\cal H}=0$ and find the analytical solution for the
same cases studied for classical dilaton cosmology.

\subsection*{Commutative Quantum Dilaton Cosmology}

\hspace{.5cm} We canonically quantize the Hamiltonian constraint of equation (\ref{ham}) to get the Wheeler-DeWitt equation $\hat{{\cal H}}\Psi=0$. Here we use the usual prescription $p_{x,y} \equiv -i \frac{\partial}{\partial x,y}$ to get
\begin{equation}
\left[~\frac{\partial^2}{\partial x^2} - \frac{\partial^2}{\partial y^2} + 2V_0 ~(x-y) \right] \Psi (x,y) = 0 ~~.
\end{equation}
Writing $\Psi(x,y)=\eta(x)\xi(y)$ we can write the variable separated equations for $x$ and $y$ as
\begin{equation}
\label{dif}
\frac{\partial^2 \eta}{\partial x^2} + (2V_0~x-k)~\eta = 0 ~~~~ \text{and} ~~~~ \frac{\partial^2 \xi}{\partial y^2} + (2V_0~y-k)~\xi = 0 ~~.
\end{equation}
Here $k/2$ is the separation parameter. The solutions of (\ref{dif}) are known in terms of Airy functions $Ai(z)$ and $Bi(z)$. We will discard the
functions $Bi(z)$ as the functions diverge for large values of $\vert z\vert$. So eigenfunctions of the Wheeler-DeWitt equation can be written as
\begin{equation}
\Psi_k(x,y) = Ai \left\{\frac{k-2V_0~x}{(2V_0)^{\frac{2}{3}}}\right\} ~ Ai \left\{\frac{k-2V_0~y}{(2V_0)^{\frac{2}{3}}}\right\} ~~.
\end{equation}
The wave function vanishes at the non singular boundary \cite{z21}. So $\Psi(0,0)=0$ yields
\begin{equation}
k_n=(2V_0)^{\frac{2}{3}} \alpha_n ~~,
\end{equation}
where $\alpha_n$ is the $n^{th}$ zero of the Airy function. The general solution of the Wheeler-DeWitt equation can now written as a superposition of the eigenfunctions
\begin{equation}
\Psi(x,y) =\sum_{n=1}^{\infty} C_n Ai \left\{\frac{k_n-2V_0~x}{(2V_0)^{\frac{2}{3}}}\right\} ~ Ai \left\{\frac{k_n-2V_0~y}{(2V_0)^{\frac{2}{3}}}\right\} ~~.
\end{equation}
\begin{figure}[htb]
\begin{tabular}{c}
\includegraphics[width=9cm,height=6cm]{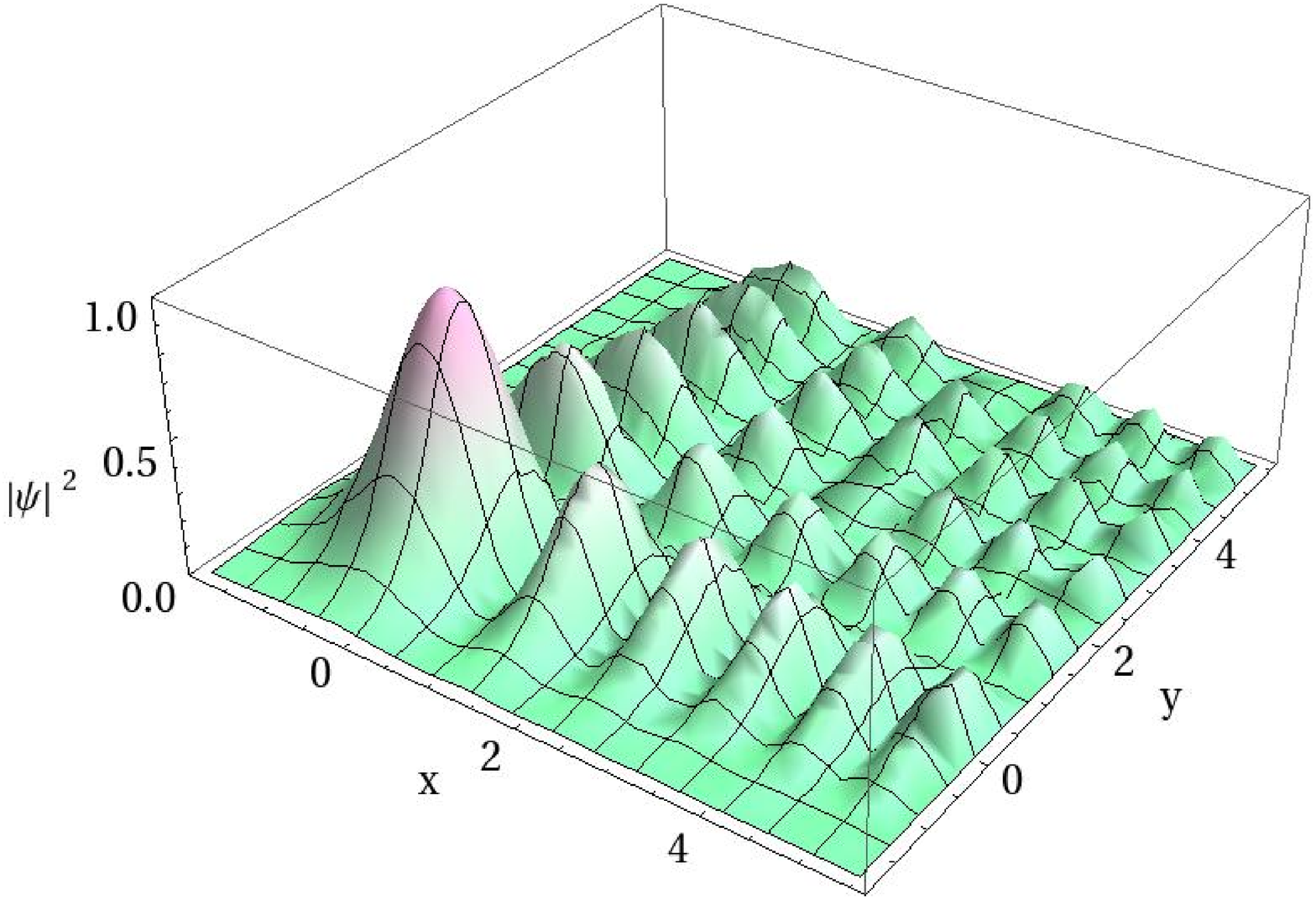} \hspace{1cm}
\includegraphics[width=5cm,height=5cm]{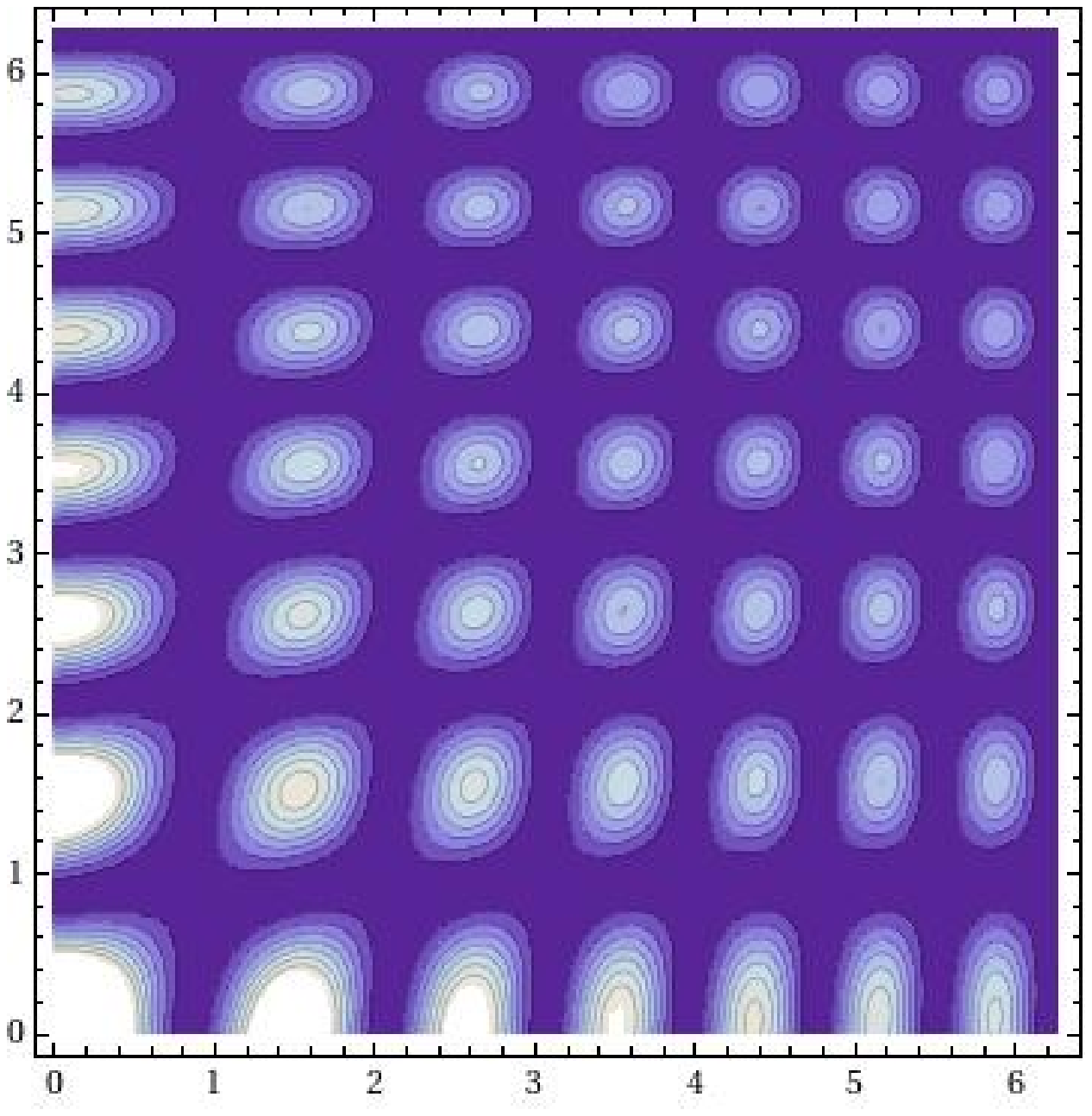}
\end{tabular}
\caption{\footnotesize  \textbf{Left} The square of the wave function in the commutative case with $V_0=1.5$. \textbf{Right} The contour plot of the same
figure on left. With other admissible values of $V_0$ the same nature of the curves are repeated. A suitable scaling is used to enlarge the figure.}
\label{psicom}
\end{figure}
In fig.(\ref{psicom}-Left) we have shown the square of the wave function in the plane of $x$ and $y$. The figure on the right shows the contour plot
of the same figure. Clearly we can see that the highest peak is around $x=0$ and $y=0$. Also there are several peaks for non-zero positive values
of $x$ and $y$. The smaller peaks are placed symmetrically around the highest peak in the first quadrant for $x$ and $y$. So we get different states but only with positive values of $x$ and $y$. This means that the dilatonic field can only have values $\phi \geq 0$ (see equation (\ref{trans})) which agrees well with the classical result for the evolution of $\phi$ (\ref{dil}). These peaks indicate that there
were different possible states from which our present universe might have evolved and tunnelled in the past. But we also see that the probability of finding
a state with zero value of $x$ is much higher.

\subsection*{Quantum Dilaton Cosmology in the MUP framework}

\hspace{.5cm} Here we will use equation (\ref{hmup}) for the Hamiltonian constraint to write the Wheeler-DeWitt equation. Following the procedure of
canonical quantization we write
\begin{equation}
\left[ \frac{\partial^2}{\partial x^2} + 2il\frac{\partial^3}{\partial x^3} - \frac{\partial^2}{\partial y^2} - 2il \frac{\partial^3}{\partial y^3}
+ 2V_0 (x-y) \right] \Psi(x,y) = 0 ~~.
\end{equation}
With the separation of variable $\Psi(x,y)=\eta(x)\xi(y)$ we can write
\begin{equation}
\label{e21}
2il~ \frac{\partial^3 \eta}{\partial x^3} + \frac{\partial^2 \eta}{\partial x^2} + (2V_0~x -k) \eta = 0 ~~,
\end{equation}
and
\begin{equation}
\label{e22}
2il~ \frac{\partial^3 \xi}{\partial y^3} + \frac{\partial^2 \xi}{\partial y^2} + (2V_0~y -k) \xi = 0 ~~,
\end{equation}
where $k/2$ is the separation parameter. Analytical closed form solution cannot be found for these equations. Here we use an approximation
method (as used in \cite{z14}) for the solution. As we are now studying the quantum model so we require the solutions in the
limit $x\rightarrow 0$ and $y\rightarrow 0$. This is physically relevant as $l$ is basically related to the Plank length. So we clearly state that our
approximation is valid only for the early universe description of the model. In the limit $l\rightarrow 0$ we get back the same equation (\ref{dif}) which is expected. So the solution remains the same i.e., Airy function. Here we expand the Airy function in the asymptotic limit $x,y \rightarrow 0$. For
the solution of equation (\ref{e21}) this can be done as
\begin{equation}
\eta(x) \sim Ai\left\{\frac{k - 2V_0~x}{(2V_0)^{\frac{2}{3}}}\right\} \sim c_0 + c_1x + c_2x^2 + c_3x^3 + \ldots ~~.
\end{equation}
Approximately in the small $x$ limit we get $\frac{\partial^3 \eta}{\partial x^3} = 6c_3$. We now put this value ($12i~l~c_3$) back in
equation (\ref{e21}) to incorporate the effect of $l$. So
\begin{equation}
\label{e23}
2il~ \frac{\partial^3 \eta}{\partial x^3} + \frac{\partial^2 \eta}{\partial x^2} + (2V_0~x -k) \eta = 12i~l~c_3 ~~.
\end{equation}
Similarly we rewrite equation (\ref{e22}) as
\begin{equation}
\label{e24}
2il~ \frac{\partial^3 \xi}{\partial y^3} + \frac{\partial^2 \xi}{\partial y^2} + (2V_0~y -k) \xi = 12i~l~c_3 ~~.
\end{equation}
The solutions of (\ref{e23}) and (\ref{e24}) can be written in terms of Airy and Hyper-geometric functions. For (\ref{e23}) we write the solution as 
\begin{equation}
\label{e29}
\eta(x) = Ai\left\{\frac{k - 2V_0~x}{(2V_0)^{\frac{2}{3}}}\right\} -i~k~{\cal C}~Ai\left\{\frac{k - 2V_0~x}{(2V_0)^{\frac{2}{3}}}\right\}~ _1{\cal F}_2 
\left\{\frac{1}{3};\frac{2}{3},\frac{4}{3};\frac{(k - 2V_0~x)^3}{36 V_0^2}\right\} + \ldots ~~,
\end{equation}
where 
\begin{equation}
{\cal C} = \frac{2\pi ~l~c_3 ~\Gamma (\frac{1}{3})}{2^{\frac{1}{3}}~3^{\frac{1}{6}} ~V_0^{\frac{4}{3}} \Gamma (\frac{2}{3}) \Gamma(\frac{4}{3})} ~~.
\end{equation}
We have neglected other terms of the solution including $Bi(z)$ as in the limit $l\rightarrow 0$ we have to get back the solution for commutative case.
If we replace $x$ by $y$ in equation (\ref{e29}) we get the solution of (\ref{e24}). Now we write the general solution of the Wheeler-DeWitt equation in the
MUP framework as
\begin{align}
\Psi_{MUP}(x,y) =& \sum_{n=1}^{\infty} C_n~\left[ Ai\left\{\frac{k_n - 2V_0~x}{(2V_0)^{\frac{2}{3}}}\right\} -i~k_n~{\cal C}~Ai\left\{\frac{k_n - 2V_0~x}{(2V_0)^{\frac{2}{3}}}\right\}~ _1{\cal F}_2 \left\{\frac{1}{3};\frac{2}{3},\frac{4}{3};\frac{(k_n - 2V_0~x)^3}{36 V_0^2}\right\} \right] \nonumber \\
& \times \left[Ai\left\{\frac{k_n - 2V_0~y}{(2V_0)^{\frac{2}{3}}}\right\} -i~k_n~{\cal C}~Ai\left\{\frac{k_n - 2V_0~y}{(2V_0)^{\frac{2}{3}}}\right\}~ _1{\cal F}_2 
\left\{\frac{1}{3};\frac{2}{3},\frac{4}{3};\frac{(k_n - 2V_0~y)^3}{36 V_0^2}\right\} \right] ~~,
\end{align}
where $k_n'^{s}$ are related to the $n^{th}$ zeros of the equation
\begin{equation}
Ai\left\{\frac{k}{(2V_0)^{\frac{2}{3}}}\right\} -i~k~{\cal C}~Ai\left\{\frac{k}{(2V_0)^{\frac{2}{3}}}\right\}~ _1{\cal F}_2 
\left\{\frac{1}{3};\frac{2}{3},\frac{4}{3};\frac{k^3}{36 V_0^2}\right\} = 0 ~~.
\end{equation}
\begin{figure}[htb]
\begin{tabular}{c}
\includegraphics[width=9cm,height=6cm]{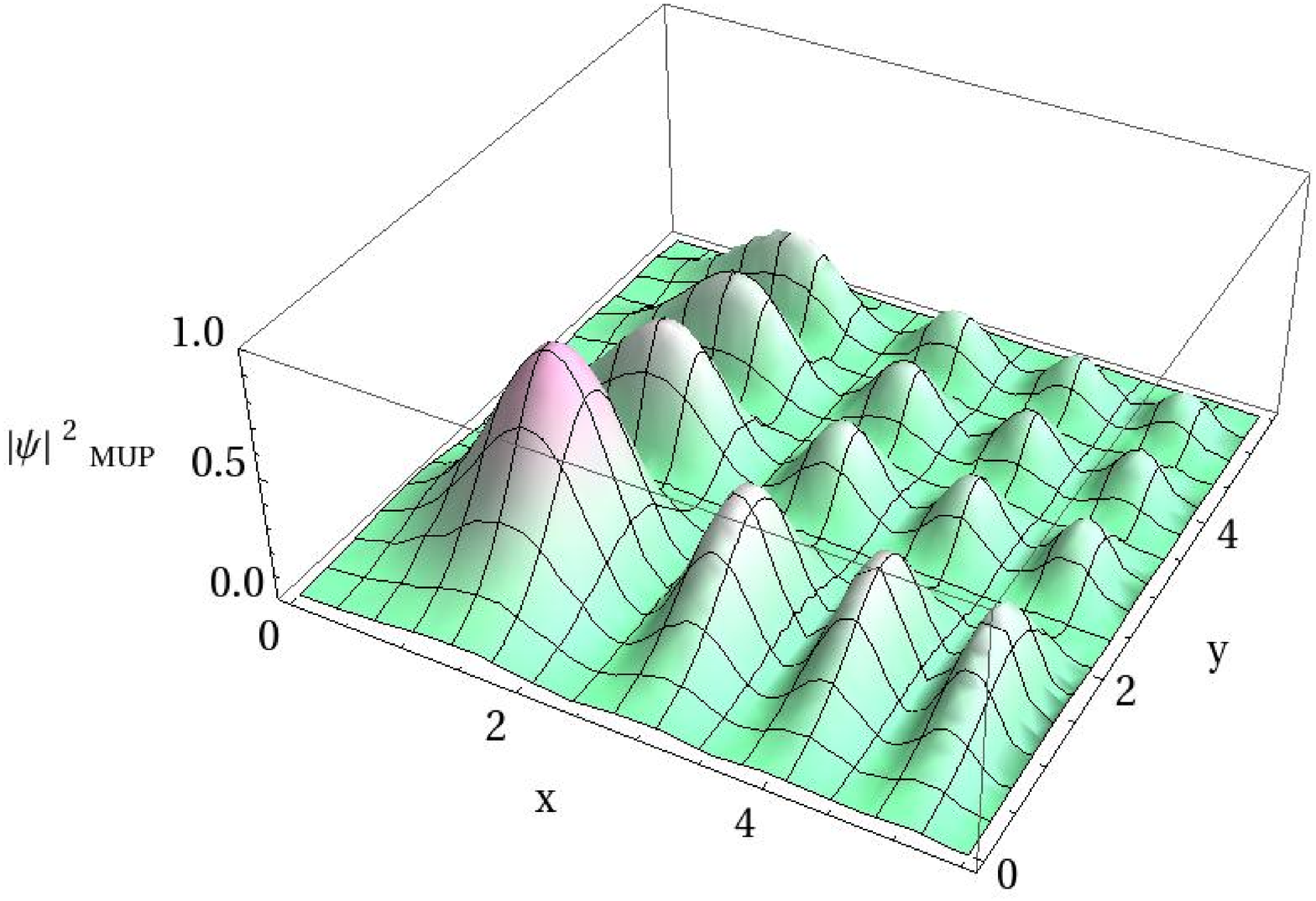} \hspace{1cm}
\includegraphics[width=5cm,height=5cm]{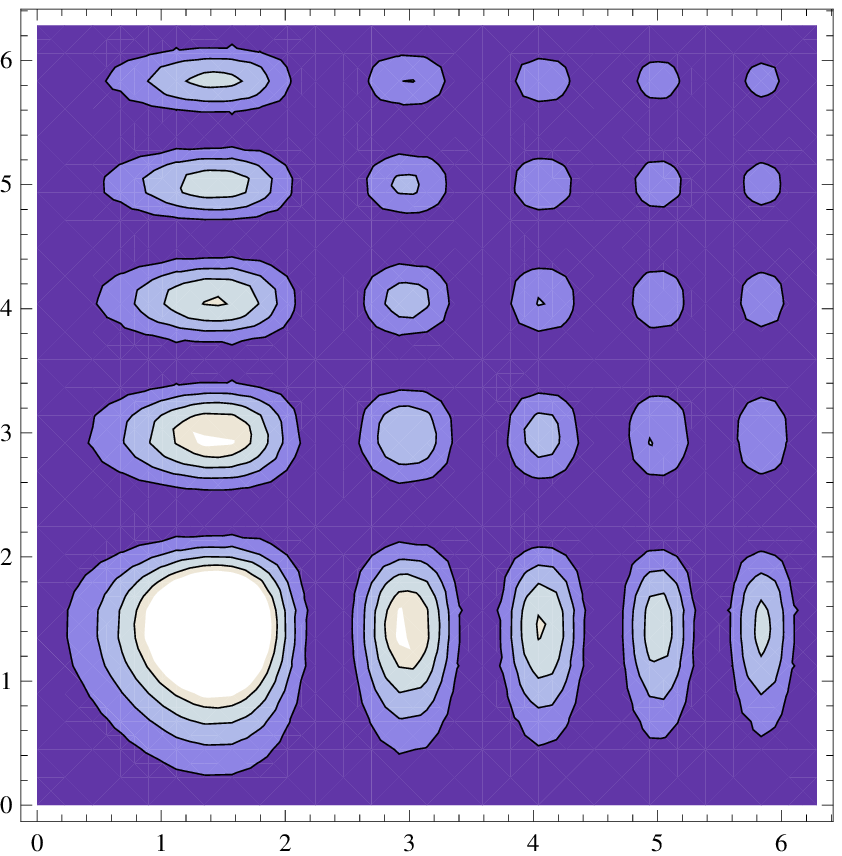}
\end{tabular}
\caption{\footnotesize  \textbf{Left} The square of the wave function in the MUP framework with $V_0=1.5$. \textbf{Right} The contour plot of the same
figure on left. With other admissible values of $V_0$ the same nature of the curves are repeated. A suitable scaling is used to enlarge the figure.}
\label{psimup}
\end{figure}
In fig.(\ref{psimup}) we have plotted the square of the MUP wave function on the $x-y$ plane for the asymptotic limit $x,y\rightarrow 0$. From the figure we can
clearly see that the largest peak is centered around non-zero values of $x$ and $y$. But in the commutative case the highest peak is around zero values
of $x$ and $y$. This difference in the position of the highest peak seems to portray the MUP effect. Here also we found several other peaks but the density is
noticeably lower than the commutative case. So possible states are less in the MUP framework.

\section{Conclusion}
 
\hspace{.5cm} In this paper we have studied the dilaton cosmological model with an exponential dilaton potential in the framework of the recently proposed modified uncertainty principle (or MUP) \cite{z15}. This MUP has a linear term in Plank length and here our aim is to study the effect of this term in the context of early universe. Replacing the commutation relation of the position and the momentum operator by their Poisson brackets we studied the classical dilaton model in the flat FRW background. As an effect of the MUP we found that the universe can undergo a late time acceleration under certain choice of some parameters of the problem. This
seems quite interesting that an extra term which modifies the usual Heisenberg algebra having its origin in the realm of quantum gravity also contributing
enormously to the late time acceleration of the universe. Later we studied quantum cosmology under the same considerations taken before. In the commutative case
we found the highest peak of the wave function is centered around $x=0$ and $y=0$. Also there are several other smaller peaks with non-zero positive values of $x$ and $y$. This could be interpreted as our universe might have evolved from one of those peaks(analogous to states in quantum mechanics). It is possible that the present universe could have evolved from one or different states with positive values of $x$ and $y$. This also indicates that the dilatonic field $\phi $ should be zero or have positive values. This result supports our classical time evolution of the dilatonic field. In the MUP framework we have presented approximate analytical solutions of the Wheeler-DeWitt equation. In this case the wave function peaks around some non-zero positive values of $x$ and $y$. Here also we found several smaller peaks but the density of peaks is less than the commutative case.

\section*{Acknowledgements}
The author is very much thankful to Prof. Narayan Banerjee for helpful discussions and guidance. E-mail discussions with Dr. Babak Vakili is
highly acknowledged as this helped immensely in developing the manuscript. The author would like to thank an anonymous referee for enlightening comments. Helpful suggestions from Soumitra Hazra is also highly acknowledged.


\section{Appendix}


\subsection{Proof for equation (\ref{e2})}

Since Black Hole Physics and String Theory suggest a modified Heisenberg algebra (which is consistent with GUP) quadratic in the
momenta (see e.g. \cite{z15}) while DSR theories suggest one that is linear in the momenta (see e.g. \cite{z8,ext1,z9}), we try to incorporate both of the
above, and start with the most general algebra with linear and quadratic terms \cite{ext2}
\begin{equation}
\label{app1}
[x_i,p_j] = i \hbar (\delta_{ij} + \delta_{ij} \alpha_1 p+ \alpha_2 \frac{p_i p_j}{p} + \beta_1 \delta_{ij}  p^2 + \beta_2 p_i p_j) ~~.
\end{equation}
Assuming that the coordinates commute among themselves, as do the momenta, it follows from the Jacobi identity that
\begin{equation}
\label{app10}
-\left[ [x_i,x_j],p_k \right] = \left[ [x_j,p_k],x_i \right] + \left[ [p_k,x_i],x_j \right]  = 0 ~~.
\end{equation}
Employing equation (\ref{app1}) and the commutator identities, and expanding the right hand side,  we get (summation convention assumed)
\begin{eqnarray}
\label{app2}
0 &=& [[x_j,p_k],x_i]+[[p_k,x_i],x_j] \nonumber \\
&=& i\hbar (-\alpha_1 \delta_{jk} [x_i,p] - \alpha_2 [x_i,p_j p_k p^{-1}]  -  \beta_1 \delta_{jk}[x_i, p_l p_l]  -\beta_2 [x_i, p_j p_k]) - (i \leftrightarrow j) \nonumber \\
&=& i\hbar \left(-\alpha_1 \delta_{jk} [x_i,p] -\alpha_2 ([x_i,p_j] p_k p^{-1}  + p_j[x_i,p_k] p^{-1}  + p_j p_k [x_i,p^{-1}]) - \right. \nonumber \\ && \left. 
 \beta_1 \delta_{jk}\left([x_i,p_l]p_l + p_l [x_i,p_l]\right) - \beta_2 \left([x_i,p_j] p_k + p_j[x_i,p_k]\right)\right)- (i \leftrightarrow j ) ~~.
\end{eqnarray}
To simplify the right hand side of Eq.(\ref{app2}), we now evaluate the following commutators

\subsubsection{\underline{\bf $[x_i,p]$ to ${\cal O}(p)$}}

Note that
\begin{eqnarray}
[x_i,p^2] &=& [x_i, p \cdot p] = [x_i,p]p + p [x_i,p] \label{app3} \\
&=& [x_i, p_k p_k ] = [x_i,p_k]p_k + p_k [x_i, p_k] \nonumber \\
&=& i\hbar \left( \delta_{ik} + \alpha_1 p \delta_{ik} + \alpha_2 p_i p_k p^{-1} \right) p_k
+ i\hbar p_k \left( \delta_{ik}  + \alpha_1 p \delta_{ik} + \alpha_2 p_i p_k p^{-1} \right)
~~(\text{to}~{\cal O}(p)~\text{using}~(\ref{app1})) \nonumber \\
&=& 2i \hbar p_i \left[ 1 + (\alpha_1+\alpha_2) p \right] \label{app4} ~~.
\end{eqnarray}
Comparing (\ref{app3}) and (\ref{app4}), we get
\begin{equation}
\label{app5}
[x_i,p ] = i\hbar \left( p_i p^{-1} + (\alpha_1 + \alpha_2 ) p_i \right) ~~.
\end{equation}

\subsubsection{\underline{\bf $[x_i,p^{-1}]$ to ${\cal O}(p)$}}

Using
\begin{equation}
0 = [x_i,\mathbf{1}] = [x_i, p \cdot p^{-1}] = [x_i,p] p^{-1} + p [x_i,p^{-1}]
\end{equation}
it follows that
\begin{eqnarray}
[x_i, p^{-1}] &=& - p_i^{-1}[x_i,p]p^{-1} \nonumber \\
&=& -i\hbar ~p^{-1} \left(p_i p^{-1} + (\alpha_1 + \alpha_2) p_i\right) p^{-1}  \nonumber \\
&=& -i\hbar ~p_i p^{-3} \left(1 + (\alpha_1+\alpha_2) p\right)~. \label{app6}
\end{eqnarray}
Substituting (\ref{app5}) and (\ref{app6}) in (\ref{app2}) and simplifying, we get
\begin{eqnarray}
0 &=&[[x_j,p_k],x_i]+[[p_k,x_i],x_j] \nonumber \\
&=& \left((\alpha_1 - \alpha_2) p^{-1}  +  (\alpha_1^2 + 2\beta_1 - \beta_2) \right) \Delta_{jki} ~~,
\end{eqnarray}
where $\Delta_{jki} = p_i \delta_{jk} - p_j \delta_{ik}$. Thus one must have $\alpha_1=\alpha_2 \equiv -l$ (with $l>0$. The negative sign follows
from \cite{ext1}), and $\beta_2 = 2\beta_1 + \alpha_1^2$. Since from dimensional grounds it follows that $\beta \sim l^2$, for simplicity, we assume $\beta_1=l^2$. Hence $\beta_2=3l^2$, and we get equation (\ref{e2}) of this paper, namely
\begin{equation}
\label{app7}
[x_i, p_j] = i \hbar \left[  \delta_{ij} -  l  \left( p \delta_{ij} +
\frac{p_i p_j}{p} \right) + l^2  \left( p^2 \delta_{ij}  + 3 p_{i} p_{j} \right)  \right] ~~.
\end{equation}


\subsection{Discussion for definitions in equation (\ref{defi})}

We would like to express the momentum $p_j$ in terms of the
{\it low energy momentum} $p_{oj}$ (such that $[x_i,p_{oj}]=i\hbar \delta_{ij}$).
Since equation (\ref{app7}) is quadratic in $p_j$, the latter can at most be
a cubic function of the $p_{oi}$. We start with the most general form
consistent with the index structure \cite{ext2}
\begin{equation}
\label{app8}
p_j = p_{oj} + a p_o p_{oj} + b p_o^2 p_{oj} ~~,
\end{equation}
where $a \sim l$ and $b\sim l^2$. From equation (\ref{app8}) it follows that
\begin{eqnarray}
\label{app9}
[x_i, p_j] &=& [ x_i , p_{oj} + a p_o p_{oj} + b p_o^2 p_{oj}] \nonumber \\
&=& i\hbar \delta_{ij} + a\left([x_i,p_o]p_{oj} + p_o [x_i, p_{oj}]\right) + b  \left([x_i,p_o] p_o p_{oj}  + p_o[x_i,p_o] p_{oj}  + p_o^2[x_i,p_{oj}]\right) ~~.
\end{eqnarray}
Next, we use the following four results to ${\cal O}(a)$ and $[x_i,p_{oj}]=i\hbar$ in equation (\ref{app9}):

\vspace{.3cm}
\noindent
(i) $[x_i,p_o] = i \hbar~p_{oi} p_o^{-1}$, which follows from equation (\ref{app5}) when $\alpha_i =0$, as well from the corresponding Poisson bracket.

\vspace{.3cm}
\noindent
(ii) $p_j = p_{oj}(1+ap_o) + {\cal O}(a^2) \simeq p_{oj} (1 + a p)$ (from equation (\ref{app8})). Therefore $p_{oj} \simeq \frac{p_j}{1+ap} \simeq (1-ap) p_j $ ~~.

\vspace{.3cm}
\noindent
(iii) $p_o = (p_{oj} p_{oj})^{\frac{1}{2}} = \left((1-ap)^2 p_j p_j \right)^{1/2} = (1-a p) p$ ~~.

\vspace{.3cm}
\noindent
(iv) $p_{oi} p_o^{-1} p_{oj} = (1-ap) p_i (1-ap)^{-1} p^{-1} (1-ap) p_j = (1-ap) p_i p_j p^{-1}$ ~~.

\vspace{.4cm}
\noindent
Thus, equation (\ref{app10}) yields
\begin{equation}
[x_i,p_j] = i\hbar \delta_{ij} + i a\hbar \left( p \delta_{ij} + p_i p_j p^{-1} \right) + i \hbar (2b-a^2) p_i p_j + i \hbar (b-a^2) p^2 \delta_{ij} ~~.
\end{equation}
Comparing with equation (\ref{app7}), it follows that $a=-l$ and $b=2l^2$. In other words
\begin{equation}
p_j = p_{oj} - l p_o p _{oj} + 2 l^2 p_o^2 p_{oj} = p_{oj}\left( 1 - l p_o + 2 l^2 p_o^2 \right) ~~,
\end{equation}
which is equation (\ref{defi}) of this paper.



\begin{thebibliography}{100}

\bibitem{z1} M. Gasperini and G. Veneziano, Mod. Phys. Lett. A {\bf 8}, 3701 (1993);\\
M. Gasperini and G. Veneziano, Phys. Rev. D {\bf 50}, 2519 (1994).
\bibitem{z2} M. Gasperini, arXiv:gr-qc/9706037.
\bibitem{z3} J. Lidsey, D. Wands, and E. Copeland, Phys. Rep. {\bf 337}, 343 (2000);\\
G. Veneziano, arXiv:hep-th/0002094;\\
F. Quevedo, Classical Quantum Gravity {\bf 19}, 5721 (2002).
\bibitem{z4} U. H. Danielsson, Classical Quantum Gravity {\bf 22}, S1 (2005).
\bibitem{z5} G. Veneziano, Phys. Lett. B {\bf 265}, 287 (1991);\\
M. Gasperini and G. Veneziano, Astropart. Phys. {\bf 1}, 317 (1993).
\bibitem{z6} M. Gasperini, arXiv:hep-th/0702166.
\bibitem{z7} C. A. Mead, Phys. Rev. D {\bf 135}, 849 (1964).
\bibitem{z8} D. J. Gross and P. F. Mende, Nucl. Phys. B {\bf 303}, 407 (1988);\\
D. Amati, M. Ciafaloni and G. Veneziano, Phys. Lett. B {\bf 216}, 41 (1989);\\
M. Kato and Phys. Lett. B {\bf 245}, 43 (1990);\\
K. Konishi, G. Paffuti and P. Provero, Phys. Lett. B {\bf 234}, 276 (1990);\\
M. Maggiore, Phys. Lett. B {\bf 304}, 65 (1993);\\
L. G. Garay, Int. J. Mod. Phys. A {\bf 10}, 145 (1995);\\
S. de Haro, JHEP {\bf 10}, 023 (1998);\\
F. Scardigli, Phys. Lett. B {\bf 452}, 39 (1999);\\
F. Brau, J. Phys. A {\bf 32}, 7691 (1999);\\
S. Hossenfelder, M. Bleicher, S. Hofmann, J. Ruppert, S. Scherer and H. Stoecker, Phys. Lett. B {\bf 575}, 85 (2003);\\
C. Bambi and F.R. Urban, Class. Quantum Grav. {\bf 25}, 095006 (2008).
\bibitem{ext1} J. Magueijo, L. Smolin, Phys. Rev. Lett. {\bf 88}, 190403 (2002);\\
J. Magueijo and L. Smolin, Phys. Rev. D {\bf 71}, 026010 (2005).
\bibitem{z9} A. Kempf, G. Mangano and R. B. Mann, Phys. Rev. D {\bf 52}, 1108 (1995);\\
A. Kempf and G. Mangano, Phys. Rev. D {\bf 55}, 7909 (1997);\\
A. Kempf, J. Phys. A {\bf 30}, 2093 (1997).
\bibitem{z10} J.L. Cortes and J. Gamboa, Phys. Rev. D {\bf 71}, 065015 (2005).
\bibitem{z11} G. M. Hossain, V. Husain and S. S. Seahra, Class. Quantum Grav. {\bf 27}, 165013 (2010).
\bibitem{z12} M. Maggiore, Phys. Lett. B {\bf 319}, 83 (1993);\\
M. Maggiore, Phys. Rev. D {\bf 49}, 5182 (1994);\\
A. Kempf, J. Math. Phys. (N.Y.) {\bf 35}, 4483 (1994);\\
H. Hinvichsen and A. Kempf, J. Math. Phys. (N.Y.) {\bf 37}, 2121 (1996);\\
R. J. Adler and D. I. Santiago, Mod. Phys. Lett. A {\bf 14}, 1371 (1999);\\
A. Kempf, Phys. Rev. D {\bf 63}, 083514 (2001);\\
A. Kempf and J. C. Niemeyer, Phys. Rev. D {\bf 64}, 103501 (2001);\\ 
L. N. Chang, D. Minic, N. Okamura and T. Takeuchi, Phys. Rev. D {\bf 65}, 125027 (2002).
\bibitem{z13} B. Vakili and H. R. Sepangi, Phys. Lett. B {\bf 651}, 79 (2007);\\
M. V. Battisti and G. Montani, Phys. Lett. B {\bf 656}, 96 (2007);\\
M. V. Battisti and G. Montani, Phys. Rev. D {\bf 77}, 023518 (2008);\\
A. Bina, K. Atazadeh and S. Jalalzadeh, Int. J. Theor. Phys. {\bf 47}, 1354 (2008).
\bibitem{z14} B. Vakili, Phys. Rev. D {\bf 77}, 044023 (2008).
\bibitem{zzzz1} L. O. Pimentel and C. Mora, Gen. Rel. and Grav. {\bf 37}, 817 (2005).
\bibitem{z15} A. F. Ali, S. Das and E. C. Vagenas, Phys. Lett. B {\bf 678}, 497 (2009).
\bibitem{z16} S. Das and E. C. Vagenas, Can. J. Phys. {\bf 87}, 233 (2009);\\
S. Das, E. C. Vagenas and A. F. Ali, Phys. Lett. B {\bf 690}, 407 (2010);\\
P. Alberto, S. Das and E. C. Vagenas, Phys. Lett. A {\bf 375}, 1436 (2011);\\
A. F. Ali, Class. Quantum Grav. {\bf 28}, 065013 (2011);\\
B. Majumder, Phys. Lett. B {\bf 699}, 315 (2011);\\
P. Pedram, K. Nozari and S. H. Taheri, JHEP {\bf 03}, 093 (2011);\\
B. Majumder, Phys. Lett. B {\bf 701} 384 (2011);\\
B. Majumder, arXiv:1105.2425 [gr-qc], in press Astrophys Space Sci (doi:10.1007/s10509-011-0815-6);\\
B. Majumder, arXiv:1105.2428v1 [gr-qc];\\
B. Majumder, in press Phys. Lett. B (doi:10.1016/j.physletb.2011.08.026), arXiv:1106.0715v1 [gr-qc].
\bibitem{z17} M. Gasperini, J. Maharana and G. Veneziano, Nucl. Phys. B {\bf 472}, 349 (1996).
\bibitem{z18} J. J. Halliwell and J. Louko, Phys. Rev. D {\bf 39}, 2206 (1989);\\
L. G. Garay, J. J. Halliwell and G. A. Mena Marugan, Phys. Rev. D {\bf 43}, 2572 (1991).
\bibitem{z19} E. J. Copeland, M. Sami and S. Tsujikawa, Int. J. Mod. Phys. D {\bf 15}, 1753 (2006).
\bibitem{z20} B. Ratra, Phys. Rev. D {\bf 45}, 1913 (1992).
\bibitem{z21} A. Vilenkin, Phys. Rev. D {\bf 37}, 888 (1988).
\bibitem{z22} N. Banerjee and S. Das, Gen. Rel. Grav. {\bf 37}, 1695 (2005).
\bibitem{z23} B. Majumder, Phys. Lett. B {\bf 697}, 101 (2011).
\bibitem{ext2} A. F. Ali, S. Das and E. C. Vagenas, Phys. Rev. D {\bf 84}, 044013 (2011).

\end{thebibliography}
\end{document}